\documentclass[twoside]{article}
\usepackage{fleqn,espcrc2,epsf}

\newcommand{\beeq}{\begin{equation}}
\newcommand{\eneq}{\end{equation}}
\newcommand{\beeqa}{\begin{eqnarray*}}
\newcommand{\eneqa}{\end{eqnarray*}}

\usepackage{graphicx}
\usepackage[figuresright]{rotating}

\newcommand{\AmS}{{\protect\the\textfont2
  A\kern-.1667em\lower.5ex\hbox{M}\kern-.125emS}}

\title{{\vspace{-6em} \normalsize
\hfill \parbox{50mm}{DESY 99-138\\}}\\[22mm]
Numerical simulations of dynamical gluinos in $SU(3)$ Yang-Mills theory:
         first results.}

\author{Alessandra Feo\address{Institut f\"ur Theoretische Physik I,
                     Universit\"at M\"unster, Wilhelm-Klemm-Str.~9, \\
                     D-48149 M\"unster, Germany}
        \thanks{Talk given by Alessandra Feo at Lattice '99, Pisa, Italy},
        Robert Kirchner\address{Deutsches Elektronen Synchrotron, DESY, \\
        Notkestr.~85, D-22603 Hamburg, Germany},
        Silke Luckmann$\rm^a$,
        Istv\'an Montvay$\rm^b$,
        Gernot M\"unster$\rm^a$,
         \\[0.5em]
        DESY-M\"unster Collaboration \\[0.5em]}

\begin{document}

\begin{abstract}
In a numerical Monte Carlo simulation of $SU(3)$ Yang-Mills theory with dynamical
gluinos we have investigated the behaviour of
the expectation value of the scalar and pseudoscalar gluino condensates in order
to determine the phase structure. Preliminary results are
presented as a function of the hopping parameter.
\vspace{1pc}
\end{abstract}

\maketitle
\section{INTRODUCTION}

In the last years there has been a great progress in the understanding of the
non-perturbative properties of supersymmetric gauge theories.
Because of their highly symmetric nature, supersymmetric quantum field theories
are best suited for analytical studies, which sometimes lead to exact
solutions \cite{sei}. The basic assumption about the non-perturbative dynamics
of supersymmetric Yang-Mills (SYM) theory is that there is confinement
and spontaneous chiral symmetry breaking \cite{kon}.


\subsection{Supersymmetric Yang-Mills theory}

Since local gauge symmetries play a very important role in nature, there is a
particular interest in supersymmetric gauge theories.
The simplest examples are SYM theories, which are supersymmetric
extensions of pure gauge theories.
We shall pay our attention to the SYM action with $N=1$, where $N$ is the
number of pairs of supersymmetry generators
$Q_{i \dot{\alpha}}$, $\overline{Q}_{i \dot{\alpha}} \; (i=1,2,\ldots, N)$.
This theory is a Yang-Mills theory with a Majorana fermion in the adjoint
representation.

The action for such a $N=1$ SYM theory with a $SU(N_c)$ gauge group is given by
\beeq
\mathcal{L} = -\frac{1}{4} F_{\mu\nu}^{a}(x) \, F_{\mu\nu}^{a}(x)
                   + \frac{1}{2} \, \overline{\Psi}^{a}(x) \gamma_{\mu}
                   \mathcal{D}_{\mu} \Psi^{a}(x) ,
\eneq
\noindent
where $\Psi^{a}(x)$ is the spinor field, and $F_{\mu\nu}^{a}(x)$ the field strength
tensor, $a \in \{1,\ldots, N_{c}^2-1\}$.

Introducing a non-zero gluino mass $m_{\tilde{g}}$ breaks supersymmetry ``softly''.
Such a mass term is
\beeq
m_{\tilde{g}}( \lambda^{\alpha} \lambda_{\alpha} + \overline{\lambda}^{\dot{\alpha}}
     \overline{\lambda}_{\dot{\alpha}}) = m_{\tilde{g}} \overline{\Psi} \Psi \; .
\eneq
\noindent
Here in the first form the Majorana-Weyl components $\lambda,\, \overline{\lambda}$
are used, in the second form the Dirac-Majorana field $\Psi$.
The Yang-Mills theory of a Majorana fermion in the adjoint representation is similar
to QCD: besides the special Majorana-feature the only difference is that the fermion
is in the adjoint representation and not in the fundamental one.
As there is only a single Majorana adjoint ``flavour'', the global chiral symmetry of $N=1$
SYM is $U(1)_{\lambda}$.

The $U(1)_{\lambda}$-symmetry is anomalous: for the corresponding axial current
$J^5_{\mu} = \overline{\Psi} \gamma_{\mu} \gamma_{5} \Psi$,
with a gauge group $SU(N_c)$, we have
\beeq
\partial^{\mu} J^5_{\mu} = \frac{N_{c} g^2}{32 \pi^2}
\epsilon^{\mu\nu\rho\tau} F_{\mu\nu}^{a} F_{\rho\tau}^{a} \; .
\eneq
\noindent
However the anomaly leaves a $Z_{2 N_c}$ unbroken: 
this can be seen by noting that the transformations
  \beeq
    \Psi \mapsto e^{-i\varphi \gamma_{5}} \Psi , \quad
    \overline{\Psi} \mapsto \overline{\Psi} e^{-i\varphi \gamma_{5}}
  \eneq
\noindent
are equivalent to
\beeq
 m_{\tilde{g}} \mapsto m_{\tilde{g}} e^{-2 i\varphi \gamma_{5}}
 \eneq
\noindent
and
 \beeq
 \Theta_{\rm SYM} \mapsto \Theta_{\rm SYM} - 2 N_c \varphi \; ,
 \eneq
\noindent
where $\Theta_{\rm SYM}$ is the $\theta$-parameter of the gauge dynamics.
Since $\Theta_{\rm SYM}$ is periodic
with period $2 \pi$, for $m_{\tilde{g}}=0$ the $U(1)_{\lambda}$ symmetry is unbroken if
 \beeq
  \varphi = \varphi_k \equiv \frac{k \pi}{N_c} , \quad (k=0,1,\cdots,2 N_c -1) \; .
 \eneq

The discrete global chiral symmetry $Z_{2 N_c}$ is expected to be spontaneously
broken to $Z_{2}$ by the non-zero gluino condensate
$ \left\langle \overline{\Psi}(x) \Psi(x)\right\rangle \neq 0$.
The consequence of this spontaneous chiral symmetry breaking is the existence
of a first order phase transition at zero gluino mass $m_{\tilde{g}} = 0$.
In the case of $N_c=2$, there exist two degenerate ground states
with opposite signs of the gluino condensate.
An interesting point is the dependence of the phase structure on the gauge group:
instanton calculations \cite{inst} at $\Theta_{\rm SYM} = 0$ give $N_c$ degenerate
vacua $(k=0,\cdots,N_c -1)$ with
\beeq
<\overline{\lambda} \lambda > = c \Lambda^3_{\rm SYM}\; e^{\frac{2 \pi i k}{N_c}} \; .
\eneq

The coexistence of $N_c$ vacua implies a first order phase transition at
$m_{\tilde{g}}=0$.
Recently Kovner and Shifman have suggested the existence of an additional
massless phase with no chiral symmetry breaking \cite{shi}.
In the case of $SU(3)$, there are at least three degenerate vacua and
for $m_{\tilde{g}} < 0$ we expect that
$\Theta_{\rm SYM} = \pi$.


\section{LATTICE FORMULATION}

No lattice gauge theory exists with an exact supersymmetry.
This is because lacking lattice generators of the Poincar\'{e} group,
it is impossible to fulfill the (continuum) algebra of SUSY transformations.
Another problem is represented by the balancing between bosonic and fermionic modes
required by SUSY: the naive lattice fermion formulation produces too many fermions.

Curci and Veneziano \cite{curci} have proposed a simple solution: 
instead of trying to have an exact version of SUSY on the lattice, 
the requirement is that, like chiral symmetry, it should
only be recovered in the continuum limit, tuning the bare parameters
(gauge coupling $g$, gluino mass $m_{\tilde{g}}$) to the supersymmetric point.

\subsection{Actions}
The Curci-Veneziano action of $N=1$ SYM is based on Wilson fermions. The effective
action obtained after integrating the gluino field is given by
\beeq
S_{CV}= \beta \sum_{pl} (1 - \frac{1}{2} \mbox{Tr} U_{pl}) -
  \frac{1}{2} \; \mbox{log} \; \mbox{det} Q[U] \; .
\eneq
The {\em fermion matrix} for the gluino $Q$ is
\beeq
Q_{yv,xu} = \delta_{yx}\delta_{vu} - 
K\sum_{\mu=\pm} \delta_{y,x+\hat{\mu}}(1+\gamma_\mu) V_{vu,x\mu} 
\eneq
\noindent
 with the gauge link in the adjoint representation $V_{vu,x\mu}[U]=
 \mbox{Tr}(U_{x\mu}^\dagger \tau_v U_{x\mu} \tau_u )$.

\subsection{Monte Carlo simulation}
The renormalized gluino mass is obtained from the hopping parameter $K$ as
\beeq
m_{R {\tilde{g}}} = \frac{Z_m(a \mu)}{2 a}\left[\frac{1}{K} -
 \frac{1}{K_0} \right] \equiv Z_m(a \mu) m_{0 {\tilde{g}}} \; .
\eneq
Here $K_0 = K_0(\beta)$ gives the $\beta-$dependent position of the phase transition
and $\mu$ is the renormalization scale.
The renormalized gluino condensate is obtained by additive and multiplicative
renormalizations:
\beeqa
&&<\overline{\Psi}(x) \Psi(x) >_{R(\mu)} =  \\
&& \qquad \qquad Z(a \mu) \left[<\overline{\Psi}(x) \Psi(x) >  -
b_0(a \mu) \right] \; .
\eneqa

A first order phase transition should show up as a jump in the expectation
value of the gluino condensate at $K = K_0$. By tuning the hopping
parameter $K$ to $K_0 $ for a fixed gauge coupling $\beta$
one expects to see a two peak structure in the distribution of the gluino condensate.
By increasing the volume the tunneling between the two ground states
becomes less and less probable and at some point practically impossible.
It is possible to see this phase diagram in our simulations by
measuring the chiral and pseudo chiral gluino condensate: the order parameter of the
supersymmetry phase transition at zero gluino mass is the value of the gluino condensate
\beeq
\rho \equiv \frac{1}{\Omega} \sum_x (\overline{\Psi}(x) \Psi(x)) \; .
\eneq
\noindent
Additionally, for $K \geq K_0 (m_{\tilde{g}} \leq 0)$ a spontaneous CP-violation,
indicated by a nonvanishing pseudo condensate
$< \overline{\Psi}(x) \gamma_5 \Psi(x) > \neq 0 $, is expected.

We determine the value of $\rho$ on a gauge configuration by stochastic estimators
\beeq
\frac{1}{N_\eta} \sum_{i=1}^{N_\eta} \sum_{x y} (\overline{\eta}_{y,i}
  Q^{-1}_{y x} \eta_{x,i} ) \; .
\eneq
Outside the phase transition region the observed distribution of $\rho$
can be fitted well by a single Gaussian but in the transition region a good
fit can only be obtained with two Gaussians. For $SU(2)$ results are shown in
\cite{ult}.

The hopping parameter $K_0$, corresponding to zero gluino mass,
is indicated by a first order phase
transition which is due to the spontaneous discrete chiral symmetry breaking
$Z_6 \rightarrow Z_2$.
\begin{figure}[htb]
\vspace*{-20pt}
\epsfxsize=6.0cm
\epsfbox{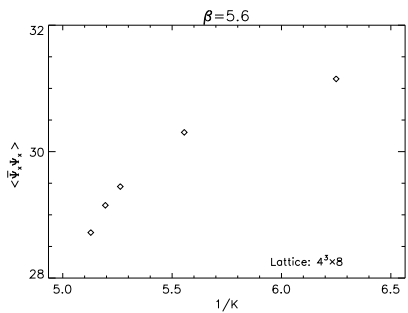}
\vspace*{-30pt}
\caption{The preliminary results for the gluino condensate.}
\label{fig01}
\vspace*{-20pt}
\end{figure}

\begin{figure}[htb]
\vspace*{-20pt}
\epsfxsize=6.0cm
\epsfbox{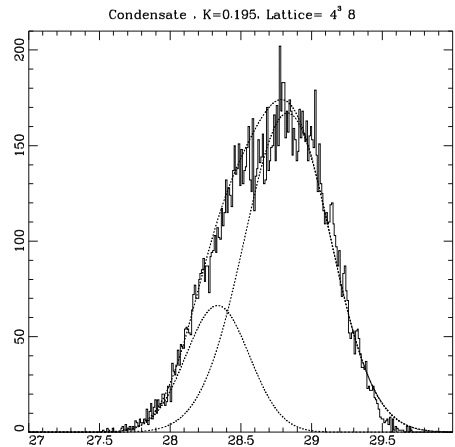}
\vspace*{-30pt}
\caption{The probability distribution of the gluino condensate for $K=0.195$ at
$\beta= 5.6$ on a $4^3 \cdot 8$ lattice.}
\vspace*{-20pt}
\label{fig02}
\end{figure}

We have investigated the dependence of the distribution of the gluino condensate
and the pseudo condensate as a function of the hopping parameter, starting from
a lattice volume $L^3 \cdot T = 4^3 \cdot 8$. This lattice is, however, still
not very large in physical units. Therefore the expected two-peak structure
is not yet very well developed, nevertheless we have high statistics.
For $K = 0.195$ fig.~\ref{fig02} shows the distribution of the gluino condensate.
The distribution indicates that we are near the phase transition.
Outside this region, we can fit the distribution with a single Gaussian.
Presently, we are calculating  on a bigger lattice volume
$(L^3 \cdot T = 6^3 \cdot 12)$ in order to separate the two-peak structure.
On the other hand, our present results on the smaller lattice do not show any signal
for a pseudo condensate.
%

{\bf Acknowledgements:}
The numerical simulations presented here have been performed on the CRAY-T3E computer
at the John von Neumann Institute for Computing (NIC) J\"ulich.
We thank NIC and the staff at ZAM for their kind support.



\end{document}